\documentclass[twoside]{elsart}
\usepackage{epsfig}
\usepackage{psfig,wrapfig,amsmath,amstext}
\journal{Physics Letters B}
\begin{document}
{
\null \hfill \begin{minipage}{4cm}IPPP/04/24 \\DCPT/04/48\\ Freiburg-THEP-04/09
\end{minipage}
}
\begin{frontmatter}
\title{On the determination of the structure of the scalar
  Higgs boson's couplings to vectorbosons}
\author{C. P. Buszello}
\address{Imperial College London, The Blackett Laboratory,
Prince Consort Road, LONDON SW7 2BW, UK}
\ead{c.buszello@imperial.ac.uk}
\author{P. Marquard}
\address{University of Durham, Department of Physics, South Road, Durham DH1 3LE, UK}
\ead{Peter.Marquard@durham.ac.uk}
\author{J. J. van der Bij}
\address{University of Freiburg, Department of Mathematics and
  Physics, Hermann-Herder-Str.10, 79104 Freiburg
i. Br., Germany}
\ead{jochum@phyv1.physik.uni-freiburg.de}
\begin{abstract}
In this paper we investigate the  determination of the coupling
structure of a  Higgs boson  at the LHC using angular correlations in the decay chain
H $\rightarrow$ ZZ $\rightarrow$ l$^+_1$l$^-_1$l$^+_2$l$^-_2$. We
consider the most general couplings of a scalar to spin 1 particles
and compare the angular correlations of the decay products using a
maximum likelihood method. We use the full information from the LO matrix element 
including all posible correlations between the decay angles.
In our analysis we  include all
possible mixings between the different coupling structures.
We conclude that the coupling structure can in general be determined using this approach.
But it has to be noted, that for Higgs boson masses below 
the ZZ-threshold the analysis is statistically limited. For higher Higgs boson masses reasonably
strong limits
on non standard 
couplings can be achieved at the LHC using the  full integrated luminosity of 300 fb$^{-1}$. 
\end{abstract}
\end{frontmatter}
\section{Introduction}
In a recently published article \cite{dist1}, see also \cite{Choi:2002jk},
 we have demonstrated that
the angular correlations of the decay leptons of Z pairs originating
from Higgs boson decays can be used to distinguish the spin/CP state $0^+$
of a Higgs boson from hypothetical states with quantum numbers $0^-$, $1^+$
and $1^-$. Only pure CP eigenstates were considered in that article.
In this letter we  show that the analysis can be generalized in order 
to achieve a more
complete determination of the coupling structure of the scalar
standard model Higgs boson.  We are using the full lagrangian from
\cite{dist1} for the coupling of a spin 0 particle to two vector bosons:
\begin{equation}
  \label{eq:0}
  {\mathcal L} = \mathbf{X} \delta^{\mu\nu} + \mathbf{Y}\frac{k^\mu
  k^\nu}{M_h^2} + \mathbf{P} \epsilon^{\mu\nu\rho\sigma}
  \frac{p_1^\rho p_2^\sigma}{M_h^2} ,
\end{equation}
where $k$ is the momentum of the Higgs boson and $p_1,p_2$ are the
momenta of the Z bosons. One or more of the parameters ${\bf X}$,
${\bf Y}$ or ${\bf P}$ can be non-zero which allows for mixed states. 
 We  demonstrate that the
values of ${\bf Y}/{\bf X}$ and ${\bf P}/{\bf X}$ can be measured
simultaneously. We determine to what precision their consistency with zero can be
established using the Atlas-Experiment.

\section{Signal selection}
In order to gain realistic results the signal selection process has to be
taken into account carefully. We use the performance of the
Atlas-detector, but in general the method is not limited to this
particular experiment.  The event generation and reconstruction
simulation has been carried out using Atlfast within the Athena
framework\cite{athena}. We use the selection criteria as set forth in
the Atlas-TDR. Two isolated leptons are required within a
pseudorapidity region of $|\eta| < 2.5$ with a transverse momentum of
more than 15 GeV. This reliably triggers the experiment. Two
additional leptons in the same pseudorapidity region with a transverse
momentum of more than
7 GeV are required. The leptons have to have appropriate flavour and
charge to be combined to two Z bosons. \\

For Higgs masses below the threshold for the production of two
on-shell Z the reducible non-resonant background from $Zb\bar{b}$ and
$t\bar{t}$ production becomes increasingly important. In order to cope
with this background, further cuts on the masses of the Z candidates and the
impact parameter are imposed. A detailed study of the effect of these
cuts can be found in \cite{lino}. The invariant mass of one of the
lepton pairs has to lie within a mass window $m_{12}$ around the Z mass,
while the mass $m_{34}$ of the other pair has to be greater than a certain
threshold.  The value of the window and the threshold are optimized
for different Higgs masses (see table \ref{tabwin}).
\begin{table}[]
\centerline{
\begin{tabular*}{14cm}{l @{\extracolsep{\fill}} c c c c c c c}
\hline \bf  Higgs mass [GeV] & 120 & 130 & 140 & 150 & 160 & 170 & 180 \\ \hline \hline
\bf Window $\mathbf{m_{12}} $ [GeV] & 20 & 15 & 15 & 10 & 10 & 6 & 6 \\ \hline 
\bf Threshold $\mathbf{m_{34}}$ [GeV] & 15 & 20 & 25 & 30 & 45 & 45 & 60 \\ \hline
\end{tabular*}
}
\caption{Mass window $m_{12}$, for the invariant mass of one lepton
  pair, and threshold $m_{34}$  
for the second pair. (See \cite{lino})
}\label{tabwin}
\end{table}
The effect of an impact parameter cut on the purity of the sample has
been studied in detail in \cite{lino}.  By applying such a cut the
reducible backgrounds can be suppressed about ten times
below the irreducible ones. Thus the contribution of these backgrounds
is expected to be negligible. The overall lepton efficiency was assumed
to be 90\% per lepton, which is rather conservative. No K-factors have
been taken into account.  The number of events for signal and
background we found  are similar to those published in
the TDR. It is worth noting that the absolute number of background
events is - within a certain range - not crucial for the analysis.
A proper normalisation of the background with data using the sidebands
is far more important.\\ 

Lacking an underlying physical theory predicting the non standard model
like couplings, $\mathbf{Y}$ and $\mathbf{P}$, we have to give a physical
interpretation for the strength of the couplings. A normalization of
the couplings is necessary in order to have a criterium to determine whether
the non-standard couplings are large or small and thereby whether the experiment
is precise or imprecise. The arbitrariness of
the mass used to normalize the mass dimension of the coupling
constants underlines this need. 
As a simple criterium we
compare the width of the states with only ${\bf P}$, ${\bf Y}$ or ${\bf X}$
couplings and require that the full width
of these states is the same for all three couplings. This yields
factors to be applied to the parameters ${\bf P}$ and ${\bf Y}$
\begin{equation}
R_P = \sqrt{\frac{\Gamma_{SM}}{\Gamma_{P}}}\,\,\,\,\, , \,\,\,\,\,\,\, R_Y = \sqrt{\frac{\Gamma_{SM}}{\Gamma_{Y}}} \\
\end{equation}
These factors are listed in table \ref{tabRpy} for various values of the
Higgs mass.  By applying these factors we redefine the
coupling parameters so that now the coupling strength is equal for
all couplings.
\begin{equation}
{\mathbf P'}= R_P\cdot {\mathbf P} \,\,\,\,\, , \,\,\,\,\,\,\,  {\mathbf Y'}=  R_Y\cdot {\mathbf Y}
\end{equation} 
It is to be noted, that this is not the only way to parametrize the couplings.
Any self consistent interpretation of the parameters is
equally valid, and corresponds in general to a linear combination of the above couplings.
In specific models for anomalous couplings a different parametrization and normalization might be
preferred.

\begin{table}\centerline{
\begin{tabular}{l c c c c c c c c c}
\hline
\bf $\mathbf{m_H}$ [GeV] & 130 & 140  & 150 & 160 & 170 & 180 & 200 & 250 & 300 \\ \hline \hline
$\bf \sqrt{\frac{\Gamma_{Y}}{\Gamma_{SM}}}$ &0.093 & 0.106  &0.116  & 0.092  &  0.106  & 0.066 & 0.102&0.284 & 0.368 \\ \hline
$\bf \sqrt{\frac{\Gamma_{P}}{\Gamma_{SM}}}$ &0.106 &0.117 &0.125 &  0.123& 0.126& 0.102& 0.146& 0.156& 0.121\\ \hline
\end{tabular}}
\caption{Ratio of the roots of the total widths of the pure states for various Higgs masses $m_H$. 
These ratios can be used to scale the constants {\bf P} and {\bf Y} such, that the
non standard model couplings are of the same strength as the standard model coupling.}
\label{tabRpy}
\end{table}

\section{Results}
\begin{figure}
\epsfig{figure=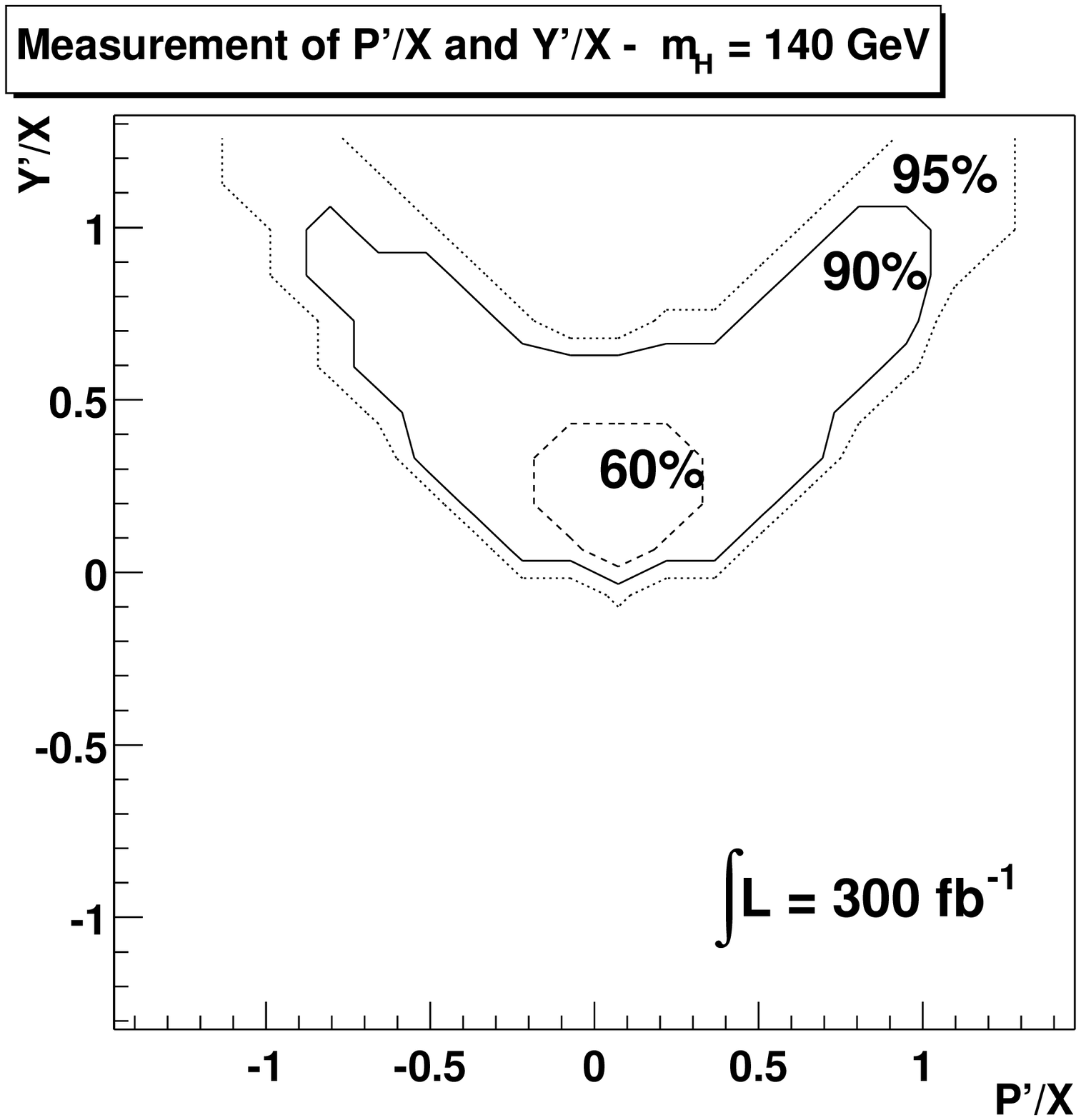,width=7.5cm}\epsfig{figure=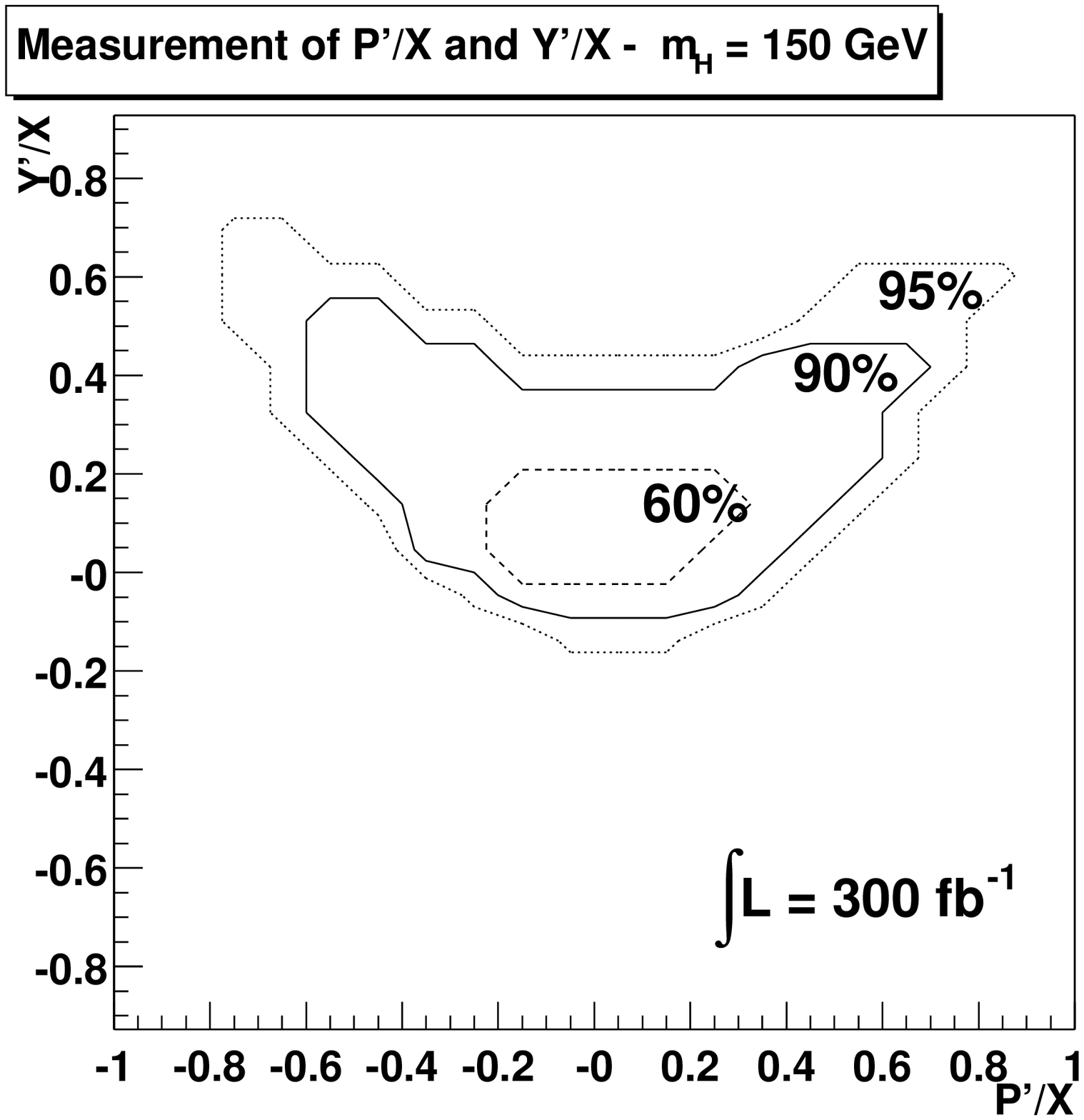,width=7.5cm}
\caption{Expected measurement of {\bf P'/X} and {\bf Y'/X} for masses of the Higgs of 140 GeV and 150 GeV.
  The quality of the measurement is mainly limited by statistics. }
\label{py150}
\end{figure}

\begin{figure}
\epsfig{figure=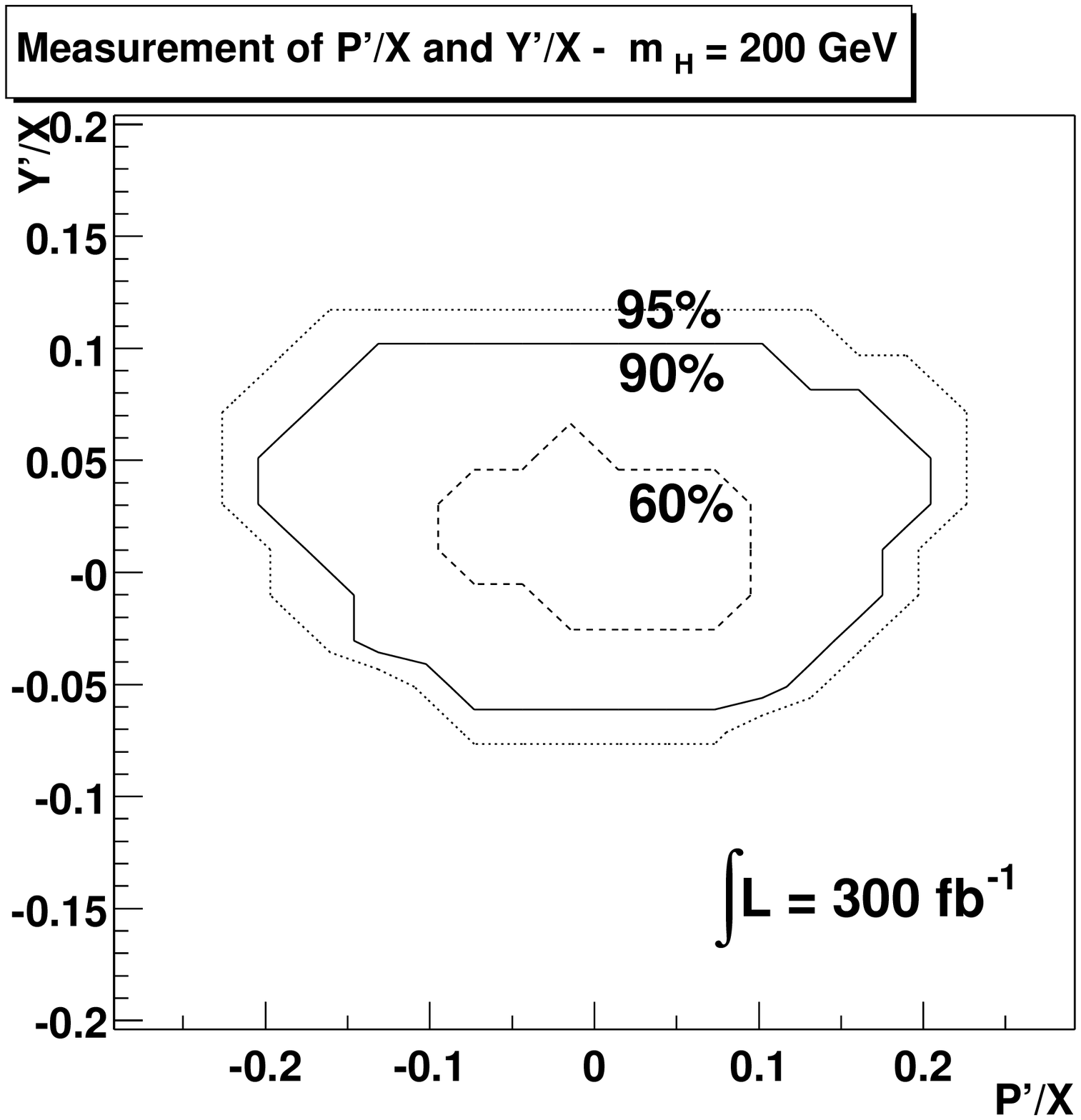,width=7.5cm}\epsfig{figure=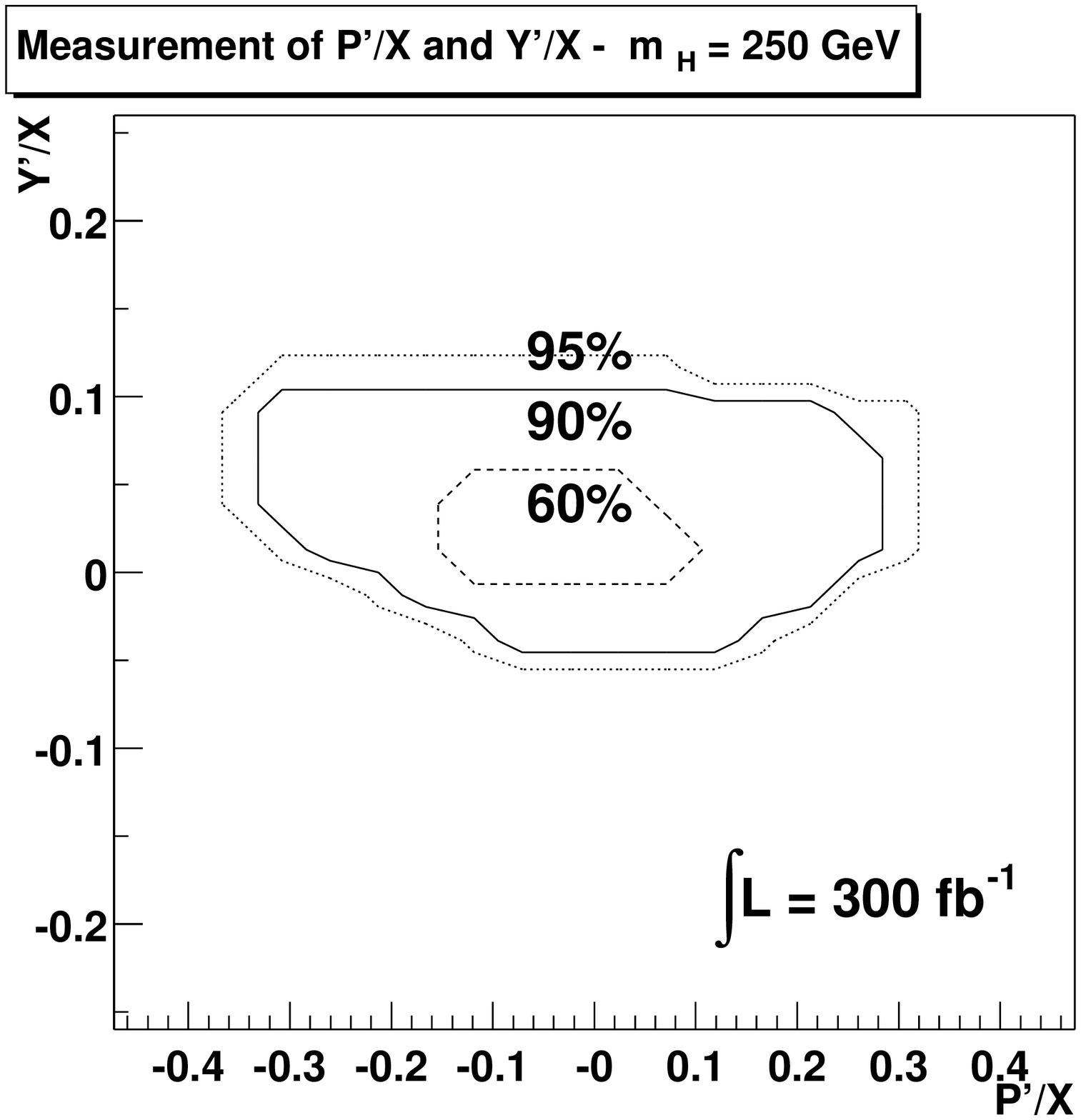,width=7.5cm}
\caption
{Expected measurement of {\bf P'/X} and {\bf Y'/X} for masses of the
  Higgs of 200 GeV and 250 GeV.  The much higher number of events
  allows for a much better measurement of the coupling structure above
  the ZZ threshold.}
\label{py200}
\end{figure}

We use the full information from the three fold differential
cross-section by constructing the following likelihood function:
\begin{equation}
 L(\mathbf{X},\mathbf{P}) = \sum_{k\in \mbox{\small events}} 
\log \frac{\mathcal{M}^2(\phi^k,\theta_1^k,\theta_2^k,\mathbf P,\mathbf Y,\mathbf  X=1)}
{ \int \mathcal{M}^2(\phi,\theta_1,\theta_2,\mathbf P,\mathbf  Y,\mathbf X=1)
 d\phi d\cos\theta_1 d\cos\theta_2}
\end{equation}
where $\mathcal{M}^2$ is the squared matrix element evaluated at
leading order.  The value of ${\bf X}$ is always fixed to the SM value
of 1, since we want to measure small contributions from non-standard
couplings. By maximising the likelihood we expect to find a value of
zero for ${\bf P'}$ and ${\bf Y'}$.  In order to demonstrate the
potential of measuring these parameters with Atlas we show
contour plots of the expected exclusion limits (see figure \ref{py150} and \ref{py200}).  The full
luminosity of 300 fb$^{-1}$ has been used for all plots. The
background has been statistically subtracted where the distribution of the 
background considered in this study was computed with Pythia.
The distortion of the signal is not negligible, but
since the contributions of the non standard model couplings are small
the distortions don't vary much. Therefore the expected likelihood
distributions are affected only slightly by the detector effects. A remarkable feature of the 
contour-plots is the V-form in the $\mathbf Y-\mathbf P$ plane. This form is understandable,
because a combination of $\mathbf Y$ and $\mathbf P$ couplings behaves very similar to
the standard model coupling $\mathbf X$.

\section{Conclusion}
\label{sec:con}
In this letter we described the analysis of the measurement of the
$HZZ$ coupling structure at the LHC using the full information of the
differential cross section ${d\sigma}/{d\cos\theta_1 d\cos\theta_2
  d\phi}$. In contrast to other studies we include possible mixings
between the different couplings.  Using the full luminosity of
\mbox{300 fb$^{-1}$} we are able to establish strong bounds on the
hypothetical coupling constants $\mathbf Y,\mathbf P$ of the order $0.1$ for a Higgs
mass above the $ZZ$ threshold. Below the threshold the results are
limited by statistics. The results are conservative, since we
worked at LO only. An inclusion of K-factors is likely to improve
the results by a factor of roughly $\sqrt{K}$. 
This analysis provides the necessary framework to perform
the measurement of the HZZ coupling structure once a Higgs boson-like
particle has been found at the LHC.

\section{Acknowledgements}
\label{sec:ack}
This work was supported in part by the BMBF, the UK Particle Physics and
Astronomy  Research Council and by the EU Fifth Framework Programme
`Improving
Human Potential', Research Training Network `Particle Physics
Phenomenology  at
High Energy Colliders', contract HPRN-CT-2000-00149.\\

This work has been performed within the ATLAS Collaboration, and we thank collaboration members for helpful discussions. We have made use of the physics analysis framework and tools which are the result of collaboration-wide efforts. 


\end{document}